\documentclass[preprint,aps,showpacs,preprintnumbers,amsmath,amssymb,nofootinbib]{revtex4-1}

\usepackage{graphicx}
\usepackage{dcolumn}
\usepackage{bm}
\usepackage{hyperref}

\begin{document}

\title{Foundations and Measures of Quantum Non-Markovianity}

\author{Heinz-Peter Breuer}
\affiliation{Physikalisches Institut, Universit\"at Freiburg,
Hermann-Herder-Strasse 3, D-79104 Freiburg, Germany}

\email[Electronic address:]{breuer@physik.uni-freiburg.de} 

\date{\today}

\begin{abstract}
The basic features of the dynamics of open quantum systems, such as 
the dissipation of energy, the decay of coherences, the relaxation to an 
equilibrium or non-equilibrium stationary state, and the transport of excitations 
in complex structures are of central importance in many applications of 
quantum mechanics. The theoretical description, analysis and control of 
non-Markovian quantum processes play an important role in this context. While in 
a Markovian process an open system irretrievably loses information to its 
surroundings, non-Markovian processes feature a flow of information from the 
environment back to the open system, which implies the presence of memory 
effects and represents the key property of non-Markovian quantum behavior. 
Here, we review recent ideas developing a general mathematical definition for 
non-Markoviantiy in the quantum regime and a measure for the degree of 
memory effects in the dynamics of open systems which are based on the 
exchange of information between system and environment. We further study the 
dynamical effects induced by the presence of system-environment correlations in 
the total initial state and design suitable methods to detect such correlations 
through local measurements on the open system.
\end{abstract}


\maketitle

\section{Introduction}

The standard approach to the theoretical description and analysis of dissipation 
and decoherence processes in open quantum systems presupposes a 
time evolution without memory. Employing the concept of a quantum Markov 
process which is given by a semigroup of completely positive dynamical maps 
one obtains a quantum Markovian master equation describing the time evolution 
of the reduced open system states with a generator in Lindblad form 
\cite{Gorini,Lindblad}. Within a microscopic approach quantum Markovian master 
equations are usually obtained by means of the Born-Markov approximation 
which assumes a weak system-environment coupling and several further, 
mostly rather drastic approximations. However, in many processes occurring in 
nature these approximations are not applicable, a situation which occurs, in 
particular, in the cases of strong system-environment couplings, structured and 
finite reservoirs, low temperatures, as well as in the presence of large initial 
system-environment correlations \cite{TheWork}. In the case of substantial 
quantitative and qualitative deviations from the dynamics of a quantum Markov 
process one often speaks of a non-Markovian process, implying that the 
dynamics is governed by significant memory effects. However, a consistent 
general theory of quantum non-Markovianity does not exist and even the very 
definition of non-Markovianity is a highly topical issue. Very recently important 
steps towards a general theory of non-Markoviantiy have been made which try to 
rigorously define the border between Markovian and non-Markovian quantum 
dynamics and to develop quantitative measures for the degree of memory effects 
in open systems \cite{Wolf,BLP,RHP}. These approaches provide general 
mathematical characterizations of quantum non-Markovianity which are 
independent from any specific representation or approximation of the dynamics, 
e.g. in terms of a perturbative master equation.

The key questions to be studied here are, how can one mathematically define and 
quantify non-Markovian behavior in the quantum regime, how do quantum 
memory effects manifest themselves in the dynamical behavior of complex open 
quantum systems, and how can such effects be uniquely identified? The answer 
to these questions is of great relevance for the design of appropriate schemes 
allowing the experimental detection and measurement of non-Markovianity. Here, 
we review some recent ideas and concepts which characterize non-Markovian 
quantum behavior by means of the information which is exchanged between an 
open quantum system and its environment \cite{BLP,MeasurePaper}. The gain or 
loss of information can be quantified through the dynamics of the trace distance 
between a pair of quantum states of the open system. It is known that this 
distance measure for quantum states can be interpreted as a measure for the 
distinguishability of the states \cite{Helstrom,Holevo,Hayashi}. 
Markovian processes tend to continuously reduce the trace distance and, hence, 
the distinguishability between any pair of physical states, which means that there 
is a flow of information from the open system to its environment. In view of this 
interpretation the characteristic feature of a non-Markovian quantum process is 
the increase of the distinguishability, i.e., a reversed flow of information from the 
environment back to the open system. Memory effects thus emerge through a 
recycling of information such that earlier states of the open system influence its 
future states.

We first recapitulate in Sec.~\ref{BASICS} all necessary concepts from the 
quantum theory of open systems, such as completely positive quantum dynamical 
maps, dynamical semigroups, Lindblad generators, quantum master equations 
and the notion of the divisibility of dynamical maps. The precise definition of 
non-Markovian quantum dynamics and the corresponding measure for the degree 
of memory effects will then be discussed in detail in Sec.~\ref{INFOFLOW}. This
section also contains a summary of the most important mathematical and
physical features of the trace distance. In Sec.~\ref{INIT-CORR} we study a
further important aspect, namely the dynamical influence of correlations in the 
initial system-environment state. We will derive general inequalities expressing
this influence and develop experimentally realizable schemes to locally detect
such correlations. Finally, some conclusions and further developments 
will be discussed in Sec.~\ref{CONCLU}.

\section{Open quantum systems: Basic notions and concepts}\label{BASICS}

We introduce and discuss some of the most important notions and 
general concepts of the quantum theory of open systems. The presentation is 
of course not an exhaustive review, but we concentrate on those topics 
that are needed for the discussions of quantum non-Markovianity in the following
sections.

\subsection{Microscopic representation of open systems}

An open quantum system $S$ is a quantum system which is coupled to another 
quantum system $E$, its environment. Thus, $S$ can be regarded as subsystem 
of the total system $S+E$ consisting of open system plus environment. 
Denoting the Hilbert spaces of $S$ and $E$ by ${\mathcal{H}}_S$ and ${\mathcal
{H}}_E$, respectively, the Hilbert space of the total system is given by the tensor 
product space 
\begin{equation}
 {\mathcal{H}} = {\mathcal{H}}_S\otimes{\mathcal{H}}_E.
\end{equation}
The physical states of the total system are described by density matrices $\rho$, 
representing positive trace class operators on ${\mathcal{H}}$ with unit
trace. This means that $\rho \geq 0$, which implies that $\rho$ is 
self-adjoint with nonnegative eigenvalues, and ${\mathrm{tr}}\rho=1$. The 
corresponding reduced states of subsystems $S$ and $E$ are given by partial 
traces over ${\mathcal{H}}_E$ and ${\mathcal{H}}_S$, respectively,
\begin{equation}
 \rho_S = {\rm{tr}}_E \rho, \qquad \rho_E = {\rm{tr}}_S \rho.
\end{equation}
In the following we denote the convex set of physical states pertaining to some 
Hilbert space ${\mathcal{H}}$ by $S({\mathcal{H}})$.

We will assume here that the total system $S+E$ is closed and follows a unitary 
dynamics described by some unitary time evolution operator
\begin{equation}
 U(t) = \exp [ -iHt ] \qquad (\hbar = 1)
\end{equation}
with a Hamiltonian of the most general form:
\begin{equation} \label{HAM-TOTAL}
 H = H_S \otimes I_E + I_S \otimes H_E + H_I,
\end{equation}
where $H_S$ and $H_E$ are the self-Hamiltonians of system and environment, respectively, and $H_I$ is any interaction Hamiltonian. Thus, the time dependence of the total system states is given by the von Neumann equation,
\begin{equation}
 \frac{d}{dt}\rho(t) = -i[H,\rho(t)],
\end{equation}
with the formal solution
\begin{equation} \label{UNITARY}
 \rho(t) = U(t)\rho(0)U^{\dagger}(t).
\end{equation}

It turns out that in most cases of practical relevance a full solution of the 
equations of motion on the microscopic level is impossible. Thus, one of the 
central goals of the quantum theory of open systems \cite{TheWork} is the 
development of an effective and analytically or numerically feasible formulation of 
the dynamical behavior of a suitably defined reduced set of variables forming the 
open subsystem $S$. Given a certain split of the total system into open system 
$S$ and environment $E$, on tries to derive effective equations of motion for the 
time dependence of the reduced system state $\rho_S(t)$ through an elimination 
of the environmental variables from the dynamical equations. The main aim is to 
develop efficient descriptions for a wide class of physical problems and 
phenomena, such as the dissipation and damping of populations, the relaxation to 
a thermal equilibrium state, the emergence of non-equilibrium stationary states,
the suppression or destruction of quantum coherences, the description of the 
quantum transport of excitations in complex systems, and the dynamics 
of quantum correlations and entanglement.

\subsection{Quantum dynamical maps} \label{DYNAMICAL-MAPS}

Quantum dynamical maps represent a key concept in the theory of open 
systems. To introduce this concept we presuppose (i) that the dynamics of the 
total system is given by a unitary time evolution (\ref{UNITARY}), and (ii) that 
system $S$ and environment $E$ are statistically independent at the initial time, 
i.e., that the initial total system state represents a tensor product state
\begin{equation} \label{PRODUCT}
 \rho(0) = \rho_S(0) \otimes \rho_E(0).
\end{equation}
While condition (i) can be relaxed to include more general cases, condition (ii) is 
crucial for the following considerations. We will discuss in Sec. \ref{INIT-CORR} 
the case of initially correlated system-environment states. On the basis of these 
assumptions the open system state at time $t \geq 0$ can be written as follows,
\begin{equation} \label{RHO-S-REPR}
 \rho_S(t) = {\rm{tr}}_E \left\{ U(t) \rho_S(0) \otimes \rho_E(0) U^{\dagger}(t) 
 \right\}.
\end{equation}
In the following we consider the initial environmental state $\rho_E(0)$ to be fixed. 
For each fixed $t \geq 0$ Eq. (\ref{RHO-S-REPR}) is then seen to represent 
a linear map 
\begin{equation} \label{MAP-PHI}
 \Phi(t,0): \, S({\mathcal{H}}_S) \longrightarrow S({\mathcal{H}}_S)
\end{equation}
on the open system's state space $S({\mathcal{H}}_S)$ which maps any initial 
open system state to the corresponding open system state at time $t$:
\begin{equation} \label{DYN-MAP}
 \rho_S(0) \mapsto  \rho_S(t) = \Phi(t,0) \rho_S(0).
\end{equation}
This is the quantum dynamical map corresponding to time $t$. It is easy to 
check that it preserves the Hermiticity and the trace of operators, i.e., we have
\begin{equation}
 {\rm{tr}}_S \left\{ \Phi(t,0) A \right\} = {\rm{tr}}_S \left\{ A \right\},
\end{equation}
and
\begin{equation}
 \left[ \Phi(t,0) A \right]^{\dagger} = \Phi(t,0)A^{\dagger}.
\end{equation}
Moreover, $\Phi(t,0)$ is a positive map, i.e., it maps positive operators to positive 
operators,
\begin{equation}
 A \geq 0 \Longrightarrow \Phi(t,0) A \geq 0.
\end{equation}
An important further property of dynamical maps is that they are not only positive 
but also completely positive. Such maps are also known as trace preserving 
quantum operations or quantum channels in quantum information and 
communication theory \cite{Nielsen}. We recall that a linear map $\Phi$ is
completely positive if and only if it admits a Kraus representation \cite{Kraus}, i.e., 
if there are operators $\Omega_i$ on the underlying Hilbert space 
${\mathcal{H}}_S$ such that
\begin{equation}
 \Phi A  = \sum_i \Omega_i A \Omega_i^{\dagger}.
\end{equation}
Such a map is trace preserving if and only if the normalization condition
\begin{equation}
 \sum_i \Omega_i^{\dagger}\Omega = I_S
\end{equation}
holds. The original definition of complete positivity, which is equivalent to the
existence of a Kraus representation, is the following. Consider for any 
$n=1,2,\ldots$ the tensor product ${\mathcal{H}}_S\otimes {\mathbb C}^n$, 
describing the Hilbert space of $S$ combined with an $n$-level system, and the 
corresponding linear tensor extension of $\Phi$ defined by 
$(\Phi\otimes I_n)(A\otimes B)= (\Phi A)\otimes B$. The map 
$\Phi\otimes I_n$ thus describes an operation which acts only on the first factor of 
the composite system, and leaves unchanged the second factor. The map $\Phi$ 
is then defined to be $n$-positive if $\Phi\otimes I_n$ is a positive map, and 
completely positive if $\Phi\otimes I_n$ is a positive map for all $n$. We note that 
positivity is equivalent to $1$-positivity, and that for a Hilbert space with finite 
dimension $N_S={\rm dim}\,{\mathcal{H}}_S$ complete positivity is equivalent to 
$N_S$-positivity \cite{Choi}.

If we now allow the time parameter $t$ to vary (keeping fixed the initial 
environmental state $\rho_E(0)$), we get a one-parameter family of dynamical 
maps,
\begin{equation} \label{FAMILY}
 \left\{ \Phi(t,0) \mid t \geq 0, \Phi(0,0) = I \right\},
\end{equation}
which contains the complete information on the dynamical evolution of all possible
initial system states. Thus, formally speaking a quantum process of an open 
system is given by such a one-parameter family of completely positive 
and trace preserving (CPT) quantum dynamical maps.

As a simple example we consider the decay of a two-state system into a bosonic 
reservoir \cite{TheWork,Kappi}. The total Hamiltonian of the model is given by
Eq.~\eqref{HAM-TOTAL} with the system Hamiltonian
\begin{equation} \label{H_0}
  H_S = \omega_{0}\sigma_+\sigma_-,
\end{equation}
describing a two-state system (qubit) with ground state
$|0\rangle$, excited state $|1\rangle$ and transition frequency
$\omega_0$, where $\sigma_+ = |1\rangle\langle 0|$ and
$\sigma_- = |0\rangle\langle 1|$ are the raising and lowering
operators of the qubit. The Hamiltonian of the environment,
\begin{equation} \label{H_E}
 H_E = \sum_k\omega_k b_k^\dagger b_k,
\end{equation}
represents a reservoir of harmonic oscillators with creation and annihilation 
operators $b_k^{\dagger}$ and $b_k$ satisfying
Bosonic commutation relations $[b_k,b_{k'}^{\dagger}] =
\delta_{kk'}$. The interaction Hamiltonian takes the form
\begin{equation} \label{H_I}
  H_I = \sum_k \left( g_k \sigma_+ \otimes b_k + g^*_k \sigma_- \otimes 
  b^{\dagger}_k
  \right).
\end{equation}
The model thus describes for example the coupling of the qubit to
a reservoir of electromagnetic field modes labelled by the index
$k$, with corresponding frequencies $\omega_k$ and coupling
constants $g_k$. Since we are using the rotating wave approximation
in the interaction Hamiltonian, the total number of excitations in the system,
\begin{equation}
 N = \sigma_+\sigma_- + \sum_k b_k^{\dagger}b_k,
\end{equation}
is a conserved quantity. The model therefore allows to derive an analytical
expression for the dynamical map \eqref{DYN-MAP}. Assuming the environment 
to be in the vacuum state $|0\rangle$ initially one finds:
\begin{eqnarray}
 \rho_{11}(t) &=& |G(t)|^2 \rho_{11}(0), \label{DYN-MAP-1} \\
 \rho_{00}(t) &=& \rho_{00}(0) + (1-|G(t)|^2)\rho_{11}(0), \label{DYN-MAP-2} \\
 \rho_{10}(t) &=& G(t) \rho_{10}(0), \label{DYN-MAP-3} \\
 \rho_{01}(t) &=& G^*(t) \rho_{01}(0), \label{DYN-MAP-4}
\end{eqnarray}
where the $\rho_{ij}(t)=\langle i |\rho_S(t)| j \rangle$ denote the matrix elements of 
$\rho_S(t)$. The function $G(t)$ introduced here is defined to be the solution
of the integro-differential equation
\begin{equation} \label{G-DEF}
 \frac{d}{dt}G(t) = -\int_0^t dt_1 f(t-t_1)G(t_1)
\end{equation}
corresponding to the initial condition $G(0)=1$, where the kernel $f(t-t_1)$ 
represents a certain two-point correlation function,
\begin{eqnarray}
\label{eq:7}
 f(t-t_1) &=& \langle 0|B(t)B^{\dagger}(t_1)|0\rangle e^{i\omega_0(t-t_1)}
 \nonumber \\
 &=& \sum_k |g_k|^2 e^{i(\omega_0-\omega_k)(t-t_1)}, 
\end{eqnarray}
of the environmental operators
\begin{equation}
 B(t) = \sum_k g_k b_k e^{-i \omega_k t}.
\end{equation}

These results hold for a generic environmental spectral density and the
corresponding two-point correlation function. Taking, for example, a
Lorentzian spectral density in resonance with the transition frequency 
of the qubit we find an exponential two-point correlation function
\begin{equation} \label{exp-corr}
  f(\tau) = \frac{1}{2} \gamma_0 \lambda e^{-\lambda |\tau|},
\end{equation}
where $\gamma_0$ describes the strength of the system-environment coupling 
and $\lambda$ the spectral width which is related to the environmental correlation
time by $\tau_E = \lambda^{-1}$. Solving
Eq.~\eqref{G-DEF} with this correlation function we find
\begin{equation} \label{G-JC}
 G(t) = e^{-\lambda t/2}\left[\cosh\left(\frac{dt}{2}\right)
 +\frac{\lambda}{d}\sinh\left(\frac{dt}{2} \right)\right],
\end{equation}
where $d=\sqrt{\lambda^2-2\gamma_0\lambda}$.

\subsection{Completely positive semigroups}

The simplest example of a quantum process is provided by a semigroup  of 
completely positive dynamical maps, which is often considered as prototypical 
example of a quantum Markov process. In this case one assumes that the set 
(\ref{FAMILY}) has the additional property
\begin{equation} \label{SEMIGROUP}
 \Phi(t,0) \, \Phi(s,0) = \Phi(t+s,0)
\end{equation}
for all $t,s \geq 0$ and, hence, has the structure of a semigroup. Under very 
general mathematical conditions such a semigroup has an infinitesimal generator 
${\mathcal{L}}$ which allows us to write 
\begin{equation} \label{GENERATOR}
 \Phi(t,0) = \exp [{\mathcal{L}}t]. 
\end{equation}
Accordingly, the reduced system state $\rho_S(t)$ obeys the master 
equation
\begin{equation} \label{LINDBLAD-MEQ}
 \frac{d}{dt} \rho_S(t) = \mathcal{L} \rho_S(t).
\end{equation}
The complete positivity of the semigroup leads to important statements on the
general structure of the generator. The famous 
Gorini-Kossakowski-Sudarshan-Lindblad theorem \cite{Gorini,Lindblad} states 
that ${\mathcal{L}}$ is the generator of a semigroup of completely positive 
quantum dynamical maps if and only if it has the following form,
\begin{equation} \label{LINDBLAD-GENERATOR}
 \mathcal{L}\rho_S = -i\left[H_S,\rho_S\right]
 + \sum_i {\gamma_i \left[ A_i \rho_S A_i^\dagger
 -\frac{1}{2} \left\{A_i^\dagger A_i,\rho_S\right\} \right]},
\end{equation}
where $H_S$ is a system Hamiltonian (which need not coincide with the system
Hamiltonian $H_S$ in the microscopic Hamiltonian (\ref{HAM-TOTAL})), the $A_i$ 
are arbitrary system operators, often called Lindblad operators, describing the 
various decay modes of the system, and the $\gamma_i$ are corresponding
decay rates. This theorem has many far-reaching consequences and is extremely 
useful, in particular in phenomenological approaches since it guarantees a 
time evolution which is compatible with general physical principles for any master 
equation of the above structure. On the other hand, it is in general difficult to 
justify rigorously the assumption (\ref{SEMIGROUP}) and to derive a quantum 
master equation of the form (\ref{LINDBLAD-MEQ}) starting from a given
system-environment model with a microscopic Hamiltonian (\ref{HAM-TOTAL}).
Such a derivation requires the validity of several approximations, the most
important one being the so-called Markov approximation. This approximation
presupposes a rapid decay of the two-point correlation
functions of those environmental operators that describe the system-environment 
coupling $H_I$. More precisely, if $\tau_E$ describes the temporal width of these
correlations and $\tau_R$ the relaxation or decoherence time of the system, the 
Markov approximation demands that
\begin{equation} \label{MARKOV-COND}
 \tau_E \ll \tau_R.
\end{equation}
This means that the environmental correlation time $\tau_E$ is small compared to 
the open system's relaxation or decoherence time $\tau_R$, i.e., that we have a 
separation of time scales, the environmental variables being the fast and the 
system variables being the slow variables. The Markov approximation is justified 
in many cases of physical interest. Typical examples of application are the weak 
coupling master equation, the quantum optical master equation describing the 
interaction of radiation with matter \cite{TheWork}, and the master equation for a 
test particle in a quantum gas \cite{Hornberger}. However, large 
couplings or interactions with low-temperature reservoirs can lead to strong
correlations resulting in long memory times and in a failure of
the Markov approximation. Moreover, the standard Markov condition 
(\ref{MARKOV-COND}) alone does {\textit{not}} guarantee, in general, that the 
Markovian master equation provides a reasonable description of the dynamics, 
a situation which can occur, for example, for finite and/or structured reservoirs 
\cite{BGM,Breuer2007}.

\subsection{Time-local master equations}
There exists a whole bunch of different theoretical and numerical methods 
for the treatment of open quantum systems, beyond the assumption of a 
dynamical semigroup, such as projection operator techniques 
\cite{Nakajima,Zwanzig}, influence functional and path integral techniques 
\cite{Grabert}, quantum Monte Carlo methods and stochastic 
wave function techniques \cite{Kappi,Piilo}. Here, we concentrate on a specific approach which is particularly suited for our purpose and which describes the open system dynamics in terms of a time-local master equation.

It is usually expected that the mathematical formulation of quantum
processes describing effects of finite memory times
in the system must necessarily involve equations of motion which are
non-local in time. In fact, such a description is suggested by the
Nakajima-Zwanzig projection operator technique which leads to an
integro-differential equation for the reduced density matrix
\cite{Nakajima,Zwanzig}. However, even the presence of strong
memory effects does not exclude the description of the dynamics in
terms of a quantum master equation which is local in time, as may
be seen from the following simple argument. According to Eq. (\ref{DYN-MAP})
we have $\rho_S(t) = \Phi(t,0) \rho_S(0)$. Assuming a smooth
time-dependence we may differentiate this relation to get
\begin{equation} \label{DIFF}
  \frac{d}{dt}\rho_S(t) = \dot{\Phi}(t,0) \rho_S(0),
\end{equation}
where the dot indicates the time derivative of $\Phi(t,0)$. To obtain a local master 
equation we invert the relation (\ref{DYN-MAP}), expressing $\rho_S(0)$ in terms 
of $\rho_S(t)$, which yields
\begin{equation} \label{MASTER-EQ}
 \frac{d}{dt}\rho_S(t) = \dot{\Phi}(t,0) \Phi^{-1}(t,0) \rho_S(t).
\end{equation}
Thus we see that the linear map ${\mathcal{K}}(t)=\dot{\Phi}(t,0) \Phi^{-1}(t,0)$ 
represents a time-dependent generator of the dynamics and we obtain a 
quantum master equation which is indeed local in time, providing a linear 
first-order differential equation for the open system state:
\begin{equation} \label{TCL-MEQ}
 \frac{d}{dt} \rho_S(t) = \mathcal{K}(t) \rho_S(t).
\end{equation}
We note that the above argument presupposes that the inverse
$\Phi^{-1}(t,0) $ of the map $\Phi(t,0)$ exists. It is possible that the 
inverse of $\Phi(t,0)$ and, hence, also the time-local generator $\mathcal{K}(t)$ 
do not exist. Such a situation can indeed occur for very
strong system-environment couplings (see below). However, for an analytic 
time-dependence the inverse of $\Phi(t,0)$ and the generator ${\mathcal{K}}(t)$ 
do exist apart from isolated singularities of the time axis \cite{Buzek}. It should 
also be emphasized that  $\Phi^{-1}(t,0)$ denotes the inverse of $\Phi(t,0)$ 
regarded as linear map acting on the space of operators of the reduced system. 
The important point is that this does {\textit{not}} imply that $\Phi^{-1}(t,0)$ is 
required to be completely positive. In general, the inverse map is not only not 
completely positive, but even not positive. There exists a powerful method for the 
microscopic derivation of time-local master equations of the form 
(\ref{TCL-MEQ}) which is known as time-convolutionless  projection operator 
technique \cite{TheWork,Kubo,Shibata,Royer1,Royer2,Kampen}. This technique
yields a systematic expansion of the generator of the master equation in terms of 
ordered cumulants and many examples have been treated with this method 
\cite{SpinStar,Esposito,Budini,Ferraro,Lidar}.

The generator ${\mathcal{K}}(t)$ of the time-local master equation must of course
preserve the Hermiticity and the trace, i.e., we have
\begin{equation} \label{K-HERM}
 \left[ {\mathcal{K}}(t) A \right]^{\dagger} = {\mathcal{K}}(t) A^{\dagger},
\end{equation}
and
\begin{equation} \label{K-TRACE}
 {\rm{tr}}_S \left\{ {\mathcal{K}}(t) A \right\} = 0.
\end{equation}
From these requirements it follows that the generator must be of the following 
most general form,
\begin{eqnarray} \label{TCL-GENERATOR}
 \mathcal{K}(t)\rho_S &=& -i\left[H_S(t),\rho_S\right] \nonumber \\
 && + \sum_i{\gamma_i(t)\left[A_i(t)\rho_S A_i^\dagger(t)
 -\frac{1}{2}\left\{A_i^\dagger(t)A_i(t),\rho_S\right\} \right]}. 
\end{eqnarray}
We see that the structure of the generator provides a natural 
generalization of the Lindblad structure, in which the Hamiltonian $H_S(t)$, the 
Lindblad operators $A_i(t)$ as well as the various decay rates $\gamma_i(t)$ 
may dependent on time. 

We briefly sketch the proof of (\ref{TCL-GENERATOR}) for a finite 
dimensional open system Hilbert space, ${\rm dim}\,\mathcal{H}_S=N_S$. To this 
end, we consider a fixed time $t$ and a fixed complete system of operators $F_i$
on $\mathcal{H}_S$, $i=1,2,\ldots,N_S^2$,  which are orthonormal with respect to 
the Hilbert-Schmidt scalar product, 
\begin{equation}
 {\rm tr}_S\{F_i^{\dagger}F_j\} = \delta_{ij}.
\end{equation}
Without loss of generality, we may further choose 
\begin{equation}
 F_{N_S^2}=\frac{1}{\sqrt{N_S}}I_S. 
\end{equation}
According to Lemma 2.3 of Ref. \cite{Gorini} any linear map ${\mathcal{K}}(t)$ 
satisfying Eqs.~(\ref{K-HERM}) and (\ref{K-TRACE}) can then be written as
\begin{eqnarray} 
 \mathcal{K}(t)\rho_S &=& -i\left[H_S(t),\rho_S\right] \nonumber \\
 && + \sum_{i,j=1}^{N_S^2-1}
 c_{ij}(t) \left[ F_i \rho_S F_j^{\dagger}
 -\frac{1}{2} \left\{ F_j^{\dagger} F_i , \rho_S \right\} \right] 
\end{eqnarray}
with a self-adjoint operator $H_S(t)$ and a Hermitian matrix $c(t)=(c_{ij}(t))$. 
Diagonalizing this matrix with the help of a unitary matrix $u(t)=(u_{ij}(t))$,
\begin{equation}
 u(t)c(t)u^{\dagger}(t) = 
 {\mbox{diag}}(\gamma_1(t),\gamma_2(t),\ldots,\gamma_{N_S^2-1}(t)),
\end{equation}
and introducing the new operators
\begin{equation}
 A_i(t) =  \sum_{j=1}^{N_S^2-1} u^{\ast}_{ij}(t) F_j,
\end{equation}
on obtains the form (\ref{TCL-GENERATOR}) of the generator.

The structure (\ref{TCL-GENERATOR}) takes into account the Hermiticity and 
trace preservation of the dynamics, but does not guarantee its complete positivity. 
The formulation of necessary and sufficient conditions for the complete positivity
of the dynamics of this generator is an important unsolved problem. However, in 
the case that the rates are positive for all times,
\begin{equation}
 \gamma_i(t) \geq 0,
\end{equation}
the resulting dynamics is indeed completely positive, since the
generator is then in Lindblad form for each fixed $t\geq 0$. 

As an example let us consider the dynamical map given by
Eqs.~\eqref{DYN-MAP-1}-\eqref{DYN-MAP-4}. In this case the time-local
generator takes the form \cite{Kappi}
\begin{eqnarray}
\label{TCL-GEN}
 \mathcal{K}(t)\rho_S &=& -\frac{i}{2}S(t)
 [\sigma_+\sigma_-,\rho_S] \nonumber \\
 &~& +\gamma(t)\left[ \sigma_-\rho_S\sigma_+
 -\frac{1}{2}\left\{\sigma_+\sigma_-,\rho_S\right\} \right],
 \end{eqnarray}
where we have introduced the definitions
\begin{equation} \label{eq:4}
 \gamma(t) = -2\Re\left(\frac{\dot{G}(t)}{G(t)}\right), \qquad
 S(t) = -2\Im\left(\frac{\dot{G}(t)}{G(t)}\right).
\end{equation}
With this generator Eq.~\eqref{TCL-MEQ} represents an exact master equation
of the model. The quantity $S(t)$ plays the role of a time-dependent frequency
shift, and $\gamma(t)$ can be interpreted as a time-dependent
decay rate. Due to the time dependence of these quantities the process does
generally not represent a dynamical semigroup, of course.
Note that the generator is finite as long as $G(t)\neq 0$; at the
zeros of $G(t)$ the inverse of the dynamical map $\Phi(t,0)$ does not exist. 
An example is provided by the zeros of the function of Eq.~\eqref{G-JC} 
which appear in the strong coupling regime $\gamma_0 > \lambda/2$.

We can also see explicitly how the standard Markov limit arises in this model. 
Considering the particular case \eqref{G-JC}, we observe that in the limit of small 
$\alpha=\gamma_0/\lambda$ we may approximate 
$G(t)\approx e^{-\gamma_0t/2}$. This approximation can also obtained directly 
from Eq.~\eqref{G-DEF} by replacing the two-point correlation function $f(t-t_1)$ 
with the delta-function $\gamma_0\delta(t-t_1)$, which is conventionally regarded 
as the Markovian limit. Equation \eqref{eq:4} then yields $S(t)=0$ and 
$\gamma(t)=\gamma_0$, i.e., the generator \eqref{TCL-GEN} assumes the form 
of a Lindblad generator of a quantum dynamical semigroup. The quantity $\alpha$
can also be written as the ratio of the environmental correlations time 
$\tau_E=\lambda^{-1}$ and the relaxation time $\tau_R=\gamma_0^{-1}$ of the 
system:
\begin{equation} \label{alpha}
 \alpha = \frac{\tau_E}{\tau_R}.
\end{equation}
Thus we see that the standard Markov condition \eqref{MARKOV-COND}
indeed leads to a Markovian semigroup here. We also mention that the 
time-convolutionless projection operator technique
yields an expansion of the generator \eqref{TCL-GEN} in powers of 
this ratio $\alpha$ \cite{TheWork}.

\subsection{Divisibility of dynamical maps} \label{DIVISIBILITY}

A family of dynamical maps $\Phi(t,0)$ is defined to be divisible if for
all $t_2 \geq t_1 \geq 0$ there exists a CPT map $\Phi(t_2,t_1)$ such that
the relation
\begin{equation}
 \Phi(t_2,0) = \Phi(t_2,t_1) \Phi(t_1,0).
 \label{eq:divisibility}
\end{equation}
holds. Note that in this equation the left-hand side as
well as the second factor on the right-hand side are fixed by the
given family of dynamical maps. Hence, Eq.~\eqref{eq:divisibility} requires
the existence of a certain linear transformation $\Phi(t_2,t_1)$
which maps the states at time $t_1$ to the states at time $t_2$
and represents a CPT map (which may be a unitary transformation). 
This definition differs slightly from the usual definition for the divisibility of a CPT 
map $\Lambda$, according to which $\Lambda$ is said to be divisible if there 
exist CPT maps $\Lambda_1$ and $\Lambda_2$ such that
$\Lambda=\Lambda_1\Lambda_2$, where one requires that neither
$\Lambda_1$ nor $\Lambda_2$ is a unitary transformation, for otherwise the 
statement is trivial \cite{Div}. 
There are many quantum processes which are not divisible. For instance, if 
$\Phi(t_1,0)$ is not invertible, a linear map $\Phi(t_2,t_1)$ which fulfills
Eq.~\eqref{eq:divisibility} may not exist. Moreover, even if a
linear map $\Phi(t_2,t_1)$ satisfying Eq.~\eqref{eq:divisibility}
does exist, this map needs not be completely positive, and not
even positive.

The simplest example of a divisible quantum process is given by a dynamical
semigroup. In fact, for a semigroup we have $\Phi(t,0)=\exp[\mathcal{L}t]$ 
and, hence, Eq.~\eqref{eq:divisibility} is trivially 
satisfied with the CPT map $\Phi(t_2,t_1)=\exp[\mathcal{L}(t_2-t_1)]$.

Consider now a quantum process given by the time-local master equation
(\ref{TCL-MEQ}) with a time dependent generator
(\ref{TCL-GENERATOR}). The dynamical maps can then be represented in terms
of a time-ordered exponential,
\begin{equation}
 \Phi(t,0) = {\rm T} \exp\left[\int_{0}^{t}{dt'\mathcal{K}(t')}\right], \quad t \geq 0,
 \label{eq:tp CPT}
\end{equation}
where ${\rm T}$ denotes the chronological time-ordering operator. We can also
define the maps
\begin{equation}
 \Phi(t_2,t_1) = {\rm T} \exp\left[\int_{t_1}^{t_2}{dt'\mathcal{K}(t')}\right],
 \quad t_2 \geq t_1 \geq 0,
 \label{eq:div for tp Mark}
\end{equation}
such that the composition law  $\Phi(t_2,0) = \Phi(t_2,t_1) \Phi(t_1,0)$ holds
by construction. The maps $\Phi(t_2,t_1)$ are completely positive, as is required 
by the divisibility condition \eqref{eq:divisibility}, if and only if the rates 
$\gamma_i(t)$ of the generator (\ref{TCL-GENERATOR}) are positive functions. 
Thus we see that divisibility is equivalent to positive rates in the time-local
master equation \cite{MeasurePaper}.
To prove this statement suppose first that $\gamma_i(t)\geq 0$. The generator 
$\mathcal{K}(t)$ is then in the Lindblad form for each fixed $t\geq 0$ and,
therefore, Eq.~\eqref{eq:div for tp Mark} represents completely positive maps. 
Conversely, assume that the maps defined by Eq.~\eqref{eq:div for tp Mark} are 
completely positive. It follows from this equation that the generator is given by
\begin{equation}
 {\mathcal{K}}(t) = \left.\frac{d}{d\tau}\right|_{\tau=0}
 \Phi(t+\tau,t).
\end{equation}
Since $\Phi(t+\tau,t)$ is completely positive for all $t,\tau\geq 0$ and 
satisfies $\Phi(t,t)=I$, this generator must be in Lindblad form for each fixed $t$, 
i.e., it must have the form (\ref{TCL-GENERATOR})  with $\gamma_i(t)\geq 0$.

The dynamical map $\Phi(t,0)$ given by 
Eqs.~\eqref{DYN-MAP-1}-\eqref{DYN-MAP-4} is completely positive if and only
if $|G(t)|\leq 1$. It is easy to verify that $\Phi(t,0)$ can be decomposed as in
Eq.~\eqref{eq:divisibility}, where the map $\Phi(t_2,t_1)$ is given by 
\cite{MeasurePaper}
\begin{eqnarray}
 \rho_{11}(t_2) &=& \left|\frac{G(t_2)}{G(t_1)}\right|^2 \rho_{11}(t_1), \nonumber \\
 \rho_{00}(t_2) &=& \rho_{00}(t_1)
 + \left(1-\left|\frac{G(t_2)}{G(t_1)}\right|^2\right)\rho_{11}(t_1), \nonumber \\
 \rho_{10}(t_2) &=& \frac{G(t_2)}{G(t_1)} \rho_{10}(t_1), \nonumber \\
 \rho_{01}(t_2) &=& \frac{G^*(t_2)}{G^*(t_1)} \rho_{01}(t_1).
\end{eqnarray}
It follows from these equations that a necessary and sufficient
condition for the complete positivity of $\Phi(t_2,t_1)$ is given by 
\begin{equation}
 |G(t_2)|\leq |G(t_1)|. 
\end{equation}
Thus we see that the dynamical map of
the model is divisible if and only if $|G(t)|$ is a monotonically
decreasing function of time. Note that this statement holds true
also for the case that $G(t)$ vanishes at some finite time.
The rate $\gamma(t)$ in Eq.~\eqref{eq:4} can be written as
\begin{equation}
 \gamma(t) = -\frac{2}{|G(t)|}\frac{d}{dt}|G(t)|.
\end{equation}
This shows that any increase of $|G(t)|$ leads to a negative decay rate
in the corresponding generator \eqref{TCL-GEN}, and illustrates the 
equivalence of the non-divisibility of the dynamical map and the occurrence of 
a temporarily negative rate in the time-local master equation demonstrated above.

\section{Information flow and non-Markovian quantum dynamics}
\label{INFOFLOW}

In the classical theory of stochastic processes a Markov process is defined
by the Markov condition, which is a condition for the hierarchy of the $n$-point
probability distribution functions pertaining to the process. Since such a hierarchy
does not exist in quantum mechanics the question arises, how do quantum 
memory effects manifest themselves in the dynamical behavior of complex open 
quantum systems and how one can rigorously define and quantify 
non-Markovianity in the quantum case? Of course, such
a definition must be independent of the specific mathematical representation
of the open system's dynamics, i.e., it must be formulated completely by means
of the quantum dynamical map of the process. In Ref.~\cite{RHP} two different 
concepts of non-Markovianity have been proposed. The first one uses the fact 
that local quantum operations can never increase the entanglement between the 
open system and an isomorphic auxiliary system. By employing an appropriate 
measure for entanglement, a given dynamical evolution is then defined to be 
non-Markovian if a temporary increase of the entanglement measure takes place, 
and the size of this increase provides a measure for the degree of 
non-Markovianity. Within the second concept a quantum process is
defined to be non-Markovian if and only if the dynamical map 
$\Phi(t,0)$ is non-divisible. The corresponding measure for non-Markovianity 
quantifies the amount to which the dynamical map violates the divisibility 
condition, thus representing a measure for the non-divisibility character of the 
process. 

Here, we discuss a recent approach which defines and 
quantifies the emergence of quantum memory effects entirely in terms of the 
exchange of information between the open system and its environment 
\cite{BLP}. In order to quantify this exchange of information we will use the trace distance as a measure for the distance between quantum states.

\subsection{Trace distance and distinguishability of quantum states} 
\label{TRACE-DIST}

The trace norm of a trace class operator $A$ is defined by
\begin{equation}
 ||A|| = {\rm{tr}} |A| = {\rm{tr}} \sqrt{A^{\dagger}A}.
\end{equation}
If $A$ is self-adjoint with eigenvalues $a_i$, this formula
reduces to 
\begin{equation}
 ||A|| = \sum_i |a_i|.
\end{equation}
Hence, the trace norm of a self-adjoint operator is equal to the sum of the moduli 
of its eigenvalues (counting their multiplicities). The trace norm leads to a natural 
and useful measure for the distance between two quantum states represented by 
positive operators $\rho^1$ and $\rho^2$ with unit trace, which is known as the 
trace distance:
\begin{equation}
 D(\rho^1,\rho^2) = \frac{1}{2}||\rho^1-\rho^2||
 = \frac{1}{2} {\rm{tr}}|\rho^1-\rho^2|.
 \label{eq:trdist}
\end{equation}
The trace distance is well defined and finite for all pairs of quantum states and 
provides a metric on the space $S({\mathcal{H}})$ of physical states.
We list some of the most important mathematical properties of this metric which 
will be needed later on (most of the proofs may be found in 
Refs. \cite{Hayashi,Nielsen}).

\begin{enumerate}

\item The trace distance between any pair of states satisfies
\begin{equation}
 0 \leq D(\rho^1,\rho^2) \leq 1.
\end{equation}
Of course, we have $D(\rho^1,\rho^2)=0$ if and only if $\rho^1=\rho^2$, while the 
upper bound is reached, i.e., $D(\rho^1,\rho^2)=1$ if and only if $\rho^1$ and 
$\rho^2$ are orthogonal, which means that the supports of $\rho^1$ and
$\rho^2$ are orthogonal. (The support is defined to be the space 
spanned by the eigenstates with nonzero eigenvalue.) Moreover, 
$D(\rho^1,\rho^2)$ is obviously symmetric in the input arguments, and satisfies 
the triangular inequality:
\begin{equation} \label{TRIANGLE}
 D(\rho^1,\rho^2) \leq D(\rho^1,\rho^3) + D(\rho^3,\rho^2).
\end{equation}

\item If $\rho^1=|\psi^1\rangle\langle\psi^1|$ and 
$\rho^2=|\psi^2\rangle\langle\psi^2|$ are pure states the following explicit
formula for the trace distance can easily be derived,
\begin{equation}
 D(\rho^1,\rho^2) = \sqrt{1-|\langle \psi^1 | \psi^2 \rangle|^2}.
\end{equation}
If the underlying Hilbert space is two-dimensional (qubit), spanned by basis states 
$|1\rangle$ and $|0\rangle$, the trace distance between two states with matrix
elements $\rho^1_{ij}$ and $\rho^2_{ij}$ is found to be
\begin{equation} \label{D-TWO-LEVEL}
 D(\rho^1,\rho^2) = \sqrt{a^2+|b|^2},
\end{equation}
where $a=\rho^1_{11}-\rho^2_{11}$ is the difference of the populations, and 
$b=\rho^1_{10}-\rho^2_{10}$ is the difference of the coherences of the two states.

\item The trace distance is sub-additive with respect to
tensor products of states which means that
\begin{equation} \label{SUB-1}
 D(\rho^1\otimes\sigma^1,\rho^2\otimes\sigma^2)
 \leq D(\rho^1,\rho^2) + D(\sigma^1,\sigma^2).
\end{equation}
In addition we have
\begin{equation} \label{SUB-2}
 D(\rho^1\otimes\sigma,\rho^2\otimes\sigma) = D(\rho^1,\rho^2).
\end{equation}

\item The trace distance is invariant under unitary transformations $U$,
\begin{equation} \label{UNITARY-TRAFO}
 D(U\rho^1U^{\dagger},U\rho^2U^{\dagger}) = D(\rho^1,\rho^2).
\end{equation}
More generally, all trace preserving and completely positive maps, i.e., all trace 
preserving quantum operations $\Lambda$ are contractions of the trace 
distance, 
\begin{equation} \label{CONTRACTION}
 D(\Lambda\rho^1,\Lambda\rho^2) \leq D(\rho^1,\rho^2).
\end{equation}
Note that the condition of the trace preservation is important here, and that this 
inequality also holds for the larger class of trace preserving positive 
maps \cite{Ruskai}.

\item The trace distance can be represented as the maximum of a certain 
functional:
\begin{equation} \label{MAXI}
 D(\rho^1,\rho^2) = \max_{\Pi}
 {\rm{tr}}\left\{\Pi\left(\rho^1-\rho^2\right)\right\}.
\end{equation}
The maximum is taken over all projection operators $\Pi$. Alternatively, one can 
take the maximum over all positive operators $A$ with $A \leq I$. Note that this 
formula is symmetric, i.e., we can also write 
$D(\rho^1,\rho^2) = \max_{\Pi}{\rm{tr}}\left\{\Pi\left(\rho^2-\rho^1\right)\right\}$ 
where, however, the maximum is then assumed for a different projection $\Pi$.

\end{enumerate}

The trace distance between two quantum states $\rho^1$ and $\rho^2$ has a 
direct physical interpretation which is based on the representation (\ref{MAXI}).
Consider two parties, Alice and Bob. Alice prepares a quantum 
system in one of two states $\rho^1$ or $\rho^2$ with probability $\frac{1}{2}$
each, and then sends the system to Bob. It is Bob's task to find out by a single 
measurement on the system whether the system state was $\rho^1$ or $\rho^2$. 
It turns out that Bob cannot always distinguish the states with certainty, but there 
is an optimal strategy which allows him to achieve the maximal possible success 
probability given by
\begin{equation} \label{P-MAX}
 P_{\max} = \frac{1}{2} \left[1 + D(\rho^1,\rho^2) \right].
\end{equation}
Thus we see that the trace distance represents the bias in favor of the
correct state identification which Bob can achieve through an
optimal strategy. The trace distance $D(\rho^1,\rho^2)$ can therefore
be interpreted as a measure for the distinguishability of the quantum
states $\rho^1$ and $\rho^2$ \cite{Helstrom,Holevo,Hayashi}.

According to Eq.~\eqref{MAXI}  Bob's optimal strategy consists in measuring the 
projection $\Pi$ for which the maximum in this relation is assumed, and in 
associating the outcome $\Pi=1$ with the state $\rho^1$, and the outcome 
$\Pi=0$ with the state $\rho^2$. Under the condition that the system state was 
$\rho^1$ he then has correctly identified this state with probability 
${\rm{tr}}\{\Pi\rho^1\}$, while under the condition that the system state was 
$\rho^2$ his answer is correct with probability ${\rm{tr}}\{(I-\Pi)\rho^2\}$. Since 
both possibilities occur with a probability of $\frac{1}{2}$ we obtain the success 
probability
\begin{eqnarray}
 P_{\max} &=& \frac{1}{2}{\rm{tr}}\{\Pi\rho^1\} + \frac{1}{2}{\rm{tr}}\{(I-\Pi)\rho^2\}
 \nonumber \\
 &=&  \frac{1}{2} \left[1 + {\rm{tr}}\left\{\Pi\left(\rho^1-\rho^2\right)\right\} \right] 
 \nonumber \\
 &=& \frac{1}{2} \left[1 + D(\rho^1,\rho^2) \right],
\end{eqnarray}
which proves Eq.~\eqref{P-MAX}. We further see that a state identification with 
certainty, $P_{\max}=1$, can be achieved if and only if $D(\rho^1,\rho^2)=1$, i.e., 
if and only if $\rho^1$ and $\rho^2$ are orthogonal, in which case Bob's optimal 
strategy is to measure the projection $\Pi$ onto the support of $\rho^1$ 
(or of $\rho^2$).

\subsection{Definition of non-Markovian quantum dynamics}

As we have seen in Sec.~\ref{TRACE-DIST} the trace distance 
$D(\rho^1,\rho^2)$
between two quantum states $\rho^1$ and $\rho^2$ can be interpreted as the
distinguishability of these states. An important conclusion from this interpretation
is that according to the contraction property of Eq.~(\ref{CONTRACTION}) no 
completely positive and trace preserving quantum operation can increase the 
distinguishability between quantum states. Consider two initial states 
$\rho_S^1(0)$ and $\rho_S^2(0)$ of an open system and the corresponding time 
evolved states
\begin{equation}
 \rho_S^{1,2}(t) = \Phi(t,0)\rho_S^{1,2}(0).
\end{equation}
Since the dynamical maps $\Phi(t,0)$ are completely positive and trace 
preserving, the trace distance between the time evolved 
states can never be larger than the trace distance between the initial states,
\begin{equation} \label{CONTR-DYN-MAP}
 D(\rho_S^1(t),\rho_S^2(t)) \leq  D(\rho_S^1(0),\rho_S^2(0)).
\end{equation}
Thus, no quantum process describable by a family of CPT dynamical maps can 
ever increase the distinguishability of a pair of states over its initial value. Of 
course, this general feature does not imply that  
$D(\rho_S^1(t),\rho_S^2(t))$ is a monotonically decreasing function of time. 

We can interpret the dynamical change of the distinguishability of the states of an 
open system in terms of a flow of information between the system and its
environment. When a quantum process reduces the distinguishability of
states, information is flowing from the system to the environment.
Correspondingly, an increase of the distinguishability signifies that
information flows from the environment back to the system. The
invariance under unitary transformations \eqref{UNITARY-TRAFO} 
indicates that information is preserved under the dynamics of
closed systems. On the other hand, the contraction property of
Eq.~\eqref{CONTRACTION} guarantees that the maximal amount of
information the system can recover from the environment is the
amount of information earlier flowed out the system.

Our definition for quantum non-Markovianity is based on the idea that
for Markovian processes any two quantum states become less and less 
distinguishable under the dynamics, leading to a perpetual loss of information
into the environment. Quantum memory effect thus arise if there is a temporal
flow of information from the environment to the system. The information flowing 
back from the environment allows the earlier open system states to have an effect 
on the later dynamics of the system, which implies the emergence of 
memory effects \cite{BLP}.

In view of these considerations, we thus define a quantum process described 
in terms of a family of quantum dynamical maps $\Phi(t,0)$ to be non-Markovian if 
and only if there is a pair of initial states $\rho_S^{1,2}(0)$ such that the trace 
distance between the corresponding states $\rho_S^{1,2}(t)$ increases at a certain time $t>0$:
\begin{equation} \label{eq:rate}
 \sigma(t,\rho_S^{1,2}(0)) \equiv \frac{d}{dt}D(\rho_S^1(t),\rho_S^2(t)) > 0,  
\end{equation}
where $\sigma(t,\rho_S^{1,2}(0))$ denotes the rate of change of the trace 
distance at time $t$ corresponding to the initial pair of states.

This definition for quantum non-Markovianity has many important consequences
for the general classification of quantum processes. In particular, it implies that all 
divisible families of dynamical maps are Markovian, including of course the class
of quantum dynamical semigroups. To prove
this statement suppose that $\Phi(t,0)$ is divisible. For any pair
of initial states $\rho_S^{1,2}(0)$ we then have
\begin{equation}
 \rho_S^{1,2}(t+\tau)=\Phi(t+\tau,t)\rho_S^{1,2}(t), \quad t,\tau\geq 0.
 \label{eq:div for initial}
\end{equation}
Since $\Phi(t+\tau,t)$ is a CPT map we can apply the contraction
property \eqref{CONTRACTION} to obtain:
\begin{equation}
 D(\rho_S^1(t+\tau),\rho_S^2(t+\tau))\leq D(\rho_S^1(t),\rho_S^2(t)).
 \label{eq:reduction}
\end{equation}
This shows that for all divisible dynamical maps the trace
distance decreases monotonically and that, therefore, these processes are
Markovian. 

Hence, we see that non-Markovian quantum processes must
necessarily be described by non-divisible dynamical maps and by time-local
master equations whose generator (\ref{TCL-GENERATOR}) involves
at least one temporarily negative rate $\gamma_i(t)$. However, it is important to
note that the converse of this statement is not true, i.e., there exist 
non-divisible dynamical processes which are Markovian in the sense of the above 
definition. For such processes the effect of the contribution of the decay 
channels with a negative rate is overcompensated by the channels with a positive 
decay rate, resulting in a total information flow directed from the open system to 
the environment. Further implications are discussed in \cite{Kossakowski}, and specific examples for this class of processes are constructed in
\cite{Mazzola,Haikka}.

We have interpreted the change of the distinguishability of quantum states as 
arising from an exchange of information between the open system and its 
environment. If the distinguishability decreases, for example, we say that 
information is flowing from the open system into the environment. The more 
precise meaning of this statement is that the available relative information on
the considered pair of states, that is, the information which enables one
to distinguish these states is lost from the open system, i.e., when only
measurements on the open system's degrees of freedom can be performed.
This does not imply that the information lost is now contained completely in the
reduced states of the environment, but instead, this information can also be
carried by the system-environment correlations. And vice versa, if we say that
information flows from the environment back to the system, this means that the
distinguishability of the open system states increases because information is
regained by the open system, information which was carried by the environmental
degrees of freedom or by the correlations between the degrees of freedom of 
system and environment (see also Sec.~\ref{INIT-CORR}).

\subsection{Construction of a measure for the degree of non-Markovianity}

For a non-Markovian process described by a family of dynamical maps $\Phi(t,0)$ 
information must flow from the environment to the system for some interval of 
time and thus we must have $\sigma>0$ for this time interval. A measure
of non-Markovianity should measure the total increase of the
distinguishability over the whole time evolution, i.e., the total
amount of information flowing from the environment back to the
system. This suggests defining a measure $\mathcal{N}(\Phi)$ for
the non-Markovianity of a quantum process through \cite{BLP}
\begin{equation} \label{NM-MEASURE}
 \mathcal{N}(\Phi) = \max_{\rho_S^{1,2}(0)}
 \int_{\sigma>0}{dt\,\sigma(t,\rho_S^{1,2}(0))}.
\end{equation}
The time integration is extended over all time intervals
$(a_i,b_i)$ in which $\sigma$ is positive and the maximum is taken
over all pairs of initial states. Due to Eq.~\eqref{eq:rate} the
measure can be written as
\begin{equation}
 \mathcal{N}(\Phi) =
 \max_{\rho^{1,2}(0)}\sum_i{\left[D(\rho_S^1(b_i),\rho_S^2(b_i))
 -D(\rho_S^1(a_i),\rho_S^2(a_i))\right]}.
 \label{eq:final measure}
\end{equation}
To calculate this quantity one first determines for any pair of
initial states the total growth of the trace distance over each
time interval $(a_i,b_i)$ and sums up the contribution of all
intervals. $\mathcal{N}(\Phi)$ is then obtained by determining the
maximum over all pairs of initial states. 

To illustrate our definition of non-Markovianity and the measure 
\eqref{NM-MEASURE} we again refer to the example of the dynamical
map given by Eqs.~\eqref{DYN-MAP-1}-\eqref{DYN-MAP-4}. According to
Eq.~\eqref{D-TWO-LEVEL} the time evolution of the trace distance corresponding 
to any pair of initial states $\rho_S^1(0)$ and $\rho_S^2(0)$ is given by
\begin{equation} \label{D-GT}
 D(\rho_S^1(t),\rho_S^2(t)) = |G(t)|\sqrt{|G(t)|^2a^2+|b|^2},
\end{equation}
where $a=\rho^1_{11}(0)-\rho^2_{11}(0)$ and $b=\rho^1_{10}(0)-\rho^2_{10}(0)$.
The time derivative of this expression yields
\begin{equation}
 \sigma(t,\rho_S^{1,2}(0)) = \frac{2|G(t)|^2a^2 + |b|^2}{\sqrt{|G(t)|^2a^2 + |b|^2}}
 \frac{d}{dt}|G(t)|.
 \label{eq:sigma for two-level}
\end{equation}
We conclude from this equation that the trace distance increases at time
$t$ if and only if the function $|G(t)|$ increases at this point of time. 
It follows that the process is non-Markovian, $\mathcal{N}(\Phi)>0$,
if and only if the dynamical map is non-divisible, which in turn is equivalent
to a temporarily negative rate $\gamma(t)$ in the time-local master equation
with generator \eqref{TCL-GEN}.

Consider, for example, the case of an exponential correlation
function which leads to the expression \eqref{G-JC} for the function
$G(t)$. For small couplings, $\gamma_0<\lambda/2$, this function decreases
monotonically. The dynamical map is then divisible, the rate $\gamma(t)$ is 
always positive, and the process is Markovian. However, in the strong coupling
regime, $\gamma_0>\lambda/2$, the function $|G(t)|$ starts to
oscillate, showing a non-monotonic behavior. Consequently, the
dynamical map is then no longer divisible and the process is non-Markovian.
Note that in the strong coupling regime the rate $\gamma(t)$ diverges at the 
zeros of $G(t)$. However, the master equation can still be used to describe the 
evolution between successive zeros and, therefore, the connection between 
non-Markovianity, divisibility and a negative rate in the master equation remains
valid. There is thus a threshold $\gamma_0 = \lambda/2$ for the
system-reservoir coupling below which $\mathcal{N}(\Phi)=0$. One
finds, as expected, that for $\gamma_0>\lambda/2$ the non-Markoviantiy 
increases monotonically with increasing system-environment coupling 
\cite{MeasurePaper}. Moreover, it is easy to show with the help of 
Eq.~\eqref{D-GT} that the maximum in Eq.~\eqref{NM-MEASURE} is attained for 
$a=0$ and $|b|=1$ \cite{Xu}. This means that the optimal pairs of initial states 
correspond to antipodal points on the equator of the Bloch sphere representing 
the qubit.

The non-Markovianity measure \eqref{NM-MEASURE} has been employed 
recently for the theoretical characterization and quantification of memory effects in 
various physical systems. An application to the information flow in the energy 
transfer dynamics of photosynthetic complexes has been developed in \cite
{Rebentrost}, and to ultracold atomic gases in \cite{Maniscalco}. Further 
applications to memory effects in the dynamics of qubits coupled to spin chains 
\cite{Paternostro} and to complex quantum systems with regular and chaotic 
dynamics \cite{Pineda} have been reported. A series of examples and details of 
the relation between the classical and the quantum concepts of non-Markovianity 
are discussed in \cite{NJP}.

The non-Markovianity measure \eqref{NM-MEASURE} represents a novel
experimentally accessible quantity. Quite recently, two experiments
determining this quantity have been carried out employing photons moving in 
optical fibers or birefringent materials \cite{NatPhys,EPL}. In these experiments 
the open system is provided by the polarization degrees of freedom of the 
photons, while the environment is given by their frequency (mode) degrees of 
freedom. It has been demonstrated that a complete determination of the measure 
is experimentally possible, including the realization of the maximization over initial 
states in the definition (\ref{NM-MEASURE}). It was shown further that a careful 
preparation of the initial environmental state allows to control the information flow 
between the system and its environment and to observe the transition from the 
Markovian to the non-Markovian regime through quantum state tomography 
carried out on the open system. This means that an experimental quantification 
and control of memory effects in open quantum systems is indeed feasible, which 
could be useful in the development of quantum memory and communication 
devices.

\section{The role of initial system-environment correlations} \label{INIT-CORR}

As we have seen in Sec. \ref{DYNAMICAL-MAPS} the construction of a
completely positive quantum dynamical map $\Phi(t,0)$ acting on the open 
system's state space is based on the assumption of an uncorrelated total 
initial state. In this section we will derive general inequalities which 
express the role of initial system-environment correlations in the subsequent 
dynamical evolution and discuss suitable strategies for the local detection of such
correlations \cite{Laine}.

\subsection{General relations describing the effect of initial correlations}
While in the general case of correlated initial states a quantum dynamical map 
can only be defined for a restricted set of states 
\cite{Shabani}, we can of course always introduce a linear map
\begin{equation} \label{MAP-PHI-CORR}
 \Lambda(t,0): \, S({\mathcal{H}}_S\otimes{\mathcal{H}}_E) 
 \longrightarrow S({\mathcal{H}}_S)
\end{equation}
on the total system's state space $S({\mathcal{H}}_S\otimes{\mathcal{H}}_E)$
by means of
\begin{equation} \label{DYN-MAP-CORR}
 \rho(0) \mapsto  \rho_S(t) = \Lambda(t,0) \rho(0)
 = {\rm{tr}}_E \left\{ U(t) \rho(0) U^{\dagger}(t) \right\}.
\end{equation}
This map takes any initial state $\rho(0)$ of the total system to the corresponding
reduced open system state $\rho_S(t)$ at time $t$. Since unitary 
transformations and partial traces are CPT maps, the composite maps 
$\Lambda(t,0)$ are again CPT maps. Considering a pair of total initial state
$\rho^{1,2}(0)$ and the corresponding open system states at time $t$,
\begin{equation} 
 \rho^{1,2}_S(t) = \Lambda(t,0) \rho^{1,2}(0),
\end{equation}
we then have by use of the contraction property \eqref{CONTRACTION} for CPT
maps:
\begin{equation} \label{CONTRACTION-CORR-1}
 D(\rho^1_S(t),\rho^2_S(t)) \leq D(\rho^1(0),\rho^2(0)),
\end{equation}
which means that the distinguishability of the open system states can never be
larger than the distinguishability of the total initial states. Subtracting the initial
trace distance of the open system states we obtain:
\begin{eqnarray} \label{CONTRACTION-CORR-2}
 \lefteqn{ D(\rho^1_S(t),\rho^2_S(t)) - D(\rho^1_S(0),\rho^2_S(0)) } \nonumber \\
 & \leq I(\rho^1(0),\rho^2(0)) 
 \equiv D(\rho^1(0),\rho^2(0)) - D(\rho^1_S(0),\rho^2_S(0)).
\end{eqnarray}
The increase of the trace distance between the open system states is thus
bounded from above by the quantity $I(\rho^1(0),\rho^2(0))$ on the right-hand 
side of this equation. 
This quantity represents the distinguishability of the total initial states minus
the distinguishability of the initial open system states. It can be interpreted
as the information on the total system states which is outside the open system, 
i.e., which is inaccessible through local measurements on the open system. 
Thus we see that the increase of the distinguishability of the open system states
and, hence, the flow of information to the open system is bounded by
the information which is outside the open system at the initial time.

We now show how the upper bound of Eq.~\eqref{CONTRACTION-CORR-2}
is connected to the correlations in the initial states. To this end, we use twice the
triangular inequality \eqref{TRIANGLE} to get:
\begin{eqnarray} \label{CONTRACTION-CORR-3}
 D(\rho^1(0),\rho^2(0)) &\leq& D(\rho^1(0),\rho_S^1(0)\otimes\rho_E^1(0))
 + D(\rho^2(0),\rho_S^1(0)\otimes\rho_E^1(0)) \nonumber \\
 &\leq& D(\rho^1(0),\rho_S^1(0)\otimes\rho_E^1(0))
 + D(\rho^2(0),\rho_S^2(0)\otimes\rho_E^2(0)) \nonumber \\
 && + D(\rho_S^1(0)\otimes\rho_E^1(0),\rho_S^2(0)\otimes\rho_E^2(0)).
\end{eqnarray}
With the help of the subadditivity \eqref{SUB-1} of the trace distance we find
\begin{equation} \label{CONTRACTION-CORR-4}
 D(\rho_S^1(0)\otimes\rho_E^1(0),\rho_S^2(0)\otimes\rho_E^2(0)) \leq
 D(\rho_S^1(0),\rho_S^2(0))  +  D(\rho_E^1(0),\rho_E^2(0)). 
 \end{equation}
Combining this with Eqs.~\eqref{CONTRACTION-CORR-3} and 
\eqref{CONTRACTION-CORR-2} we finally obtain
\begin{eqnarray} \label{CONTRACTION-CORR-5}
 D(\rho^1_S(t),\rho^2_S(t)) - D(\rho^1_S(0),\rho^2_S(0))
 &\leq& D(\rho^1(0),\rho_S^1(0)\otimes\rho_E^1(0))
 + D(\rho^2(0),\rho_S^2(0)\otimes\rho_E^2(0)) \nonumber \\
 && + D(\rho_E^1(0),\rho_E^2(0)). 
\end{eqnarray}
For any total system state $\rho$ the trace distance 
$D(\rho,\rho_S\otimes\rho_E)$ between $\rho$ and the product of its marginals
$\rho_S\otimes\rho_E$ represents a measure for the correlations in the state 
$\rho$, which quantifies how well $\rho$ can be distinguished from the 
corresponding product state $\rho_S\otimes\rho_E$. Thus, 
Eq.~\eqref{CONTRACTION-CORR-5} demonstrates that a dynamical 
increase of the trace distance of the open system states over the initial value,
\begin{eqnarray} \label{CONTRACTION-CORR-6}
 D(\rho^1_S(t),\rho^2_S(t)) - D(\rho^1_S(0),\rho^2_S(0)) > 0,
\end{eqnarray}
implies that the corresponding initial environmental states $\rho_E^{1,2}(0)$ are
different or that at least one of the total initial states $\rho^{1,2}(0)$ is correlated.

As an example we consider the special case of an initial pair of states
given by a correlated state $\rho^1(0)$ and the uncorrelated 
product of its marginals 
$\rho^2(0)=\rho_S^1(0)\otimes\rho_E^1(0)$. Hence, we have
$\rho_S^1(0)=\rho_S^2(0)$ and $\rho_E^1(0)=\rho_E^2(0)$, and 
Eq.~\eqref{CONTRACTION-CORR-5} reduces to
\begin{equation} \label{CONTRACTION-CORR-7}
 D(\rho^1_S(t),\rho^2_S(t)) 
 \leq D(\rho^1(0),\rho_S^1(0)\otimes\rho_E^1(0)). 
\end{equation}
Thus, the increase of the trace distance between $\rho^1_S(t)$ and
$\rho^2_S(t)$ (which is zero initially) is bounded
from above by the amount of correlations in the initial state $\rho^1(0)$.

We further remark that in the case of two uncorrelated states with the same
environmental state, i.e., $\rho^{1,2}(0)=\rho_S^{1,2}(0)\otimes\rho_E(0)$, the 
right-hand side of the inequality \eqref{CONTRACTION-CORR-5} vanishes, and
we are led again to the contraction property \eqref{CONTR-DYN-MAP} for 
CPT dynamical maps on the reduced state space. Thus, 
Eq.~\eqref{CONTRACTION-CORR-5} represents a generalization of the 
contraction property of dynamical maps on the reduced state space.

\subsection{Witnessing system-environment correlations by local operations}

Equation \eqref{CONTRACTION-CORR-5} leads to an experimentally 
realizable scheme for the local detection of correlations in an unknown
total system's initial state $\rho^1(0)$ as follows \cite{Laine}. First, one generates 
a second reference state $\rho^2(0)$ by applying a local trace-preserving 
quantum operation ${\mathcal{E}}$,
\begin{equation}
 \rho^2(0) = ({\mathcal{E}}\otimes I) \rho^1(0).
\end{equation}
The operation ${\mathcal{E}}$ acts only on the variables of the open system, and 
can be realized, for example, by the measurement of an observable of the open 
system, or by a unitary transformation induced, e.g., through an external control 
field. It is easy to check that $\rho^1(0)$ and $\rho^2(0)$ lead to the same
reduced environmental state, i.e., we have $\rho^1_E(0)=\rho^2_E(0)$ and,
hence, Eq.~\eqref{CONTRACTION-CORR-5} yields
\begin{equation} \label{CONTRACTION-CORR-8}
 D(\rho^1_S(t),\rho^2_S(t)) - D(\rho^1_S(0),\rho^2_S(0))
 \leq D(\rho^1(0),\rho_S^1(0)\otimes\rho_E^1(0))
 + D(\rho^2(0),\rho_S^2(0)\otimes\rho_E^2(0)).
\end{equation}
This inequality shows that any dynamical increase of the trace distance 
between the open system states over the initial value implies the presence of 
correlations in the initial state $\rho^1(0)$. In fact, if one finds that the left-hand 
side of the inequality is greater than zero, then either $\rho^1(0)$ or $\rho^2(0)$ 
must be correlated. If $\rho^1(0)$ was uncorrelated, then also $\rho^2(0)$ must 
be uncorrelated since it is obtained from $\rho^1(0)$ through application of a local 
operation. Thus, any increase of the trace distance of the open system states 
over the initial value witnesses correlations in $\rho^1(0)$.

We note that this strategy for the local detection of initial correlations requires
only local control and measurements of the open quantum system. It neither 
demands knowledge about the structure of the environment or of the 
system-environment interaction, nor a full knowledge of the initial
system-environment state $\rho^1(0)$. Moreover, there is no
principal restriction on the operation ${\mathcal{E}}$ used to generate the
second state $\rho^2(0)$, which makes the scheme very flexible in practice. 
In fact, experimental realizations of the scheme have been reported recently 
\cite{Hefei, Milan}. Further examples and applications to the study 
of correlations in thermal equilibrium states are discussed in Ref.~\cite{Gibbs}. 
Moreover, by taking ${\mathcal{E}}$ to be a pure dephasing operation
the scheme enables the local detection of nonclassical correlations, 
i.e., of total initial states with nonzero quantum discord \cite{Modi}. This fact has 
been shown very recently in Ref.~\cite{Manu} where also a statistical approach to 
initial correlations on the basis of random matrix theory has been developed.

\section{Conclusions}\label{CONCLU}

We have discussed a definition for the non-Markovianity of quantum processes in 
open systems and developed a corresponding measures for the size of 
quantum memory effects. Our considerations are based on the quantification of 
the information flow between the open system and its environment in terms of
the trace distance between quantum states of the open system. The great 
advantage of this distance measure is the fact that it admits a natural and clear 
physical interpretation as the distinguishability of the states through local 
measurements carried out on the open system. According to our definition the
key feature of quantum non-Markovianity is the temporal increase of the 
distinguishability which can be interpreted as a backflow of information from the
environment to the open system.  As we have seen this concept allows a natural 
extension to the case of initial system-environment correlations.
While a quantum dynamical map acting on the open system's state space does
in general not exist in this case, the trace distance between pairs of states of the 
open system leads to a dynamical witness for the presence of initial correlations
in the total system state. Given that the initial correlations were created in the past 
from a product state through a system-environment interaction, 
the increase of the trace distance of the open system states over its initial value 
signifies that the open system regains information which was lost previously.

There are of course several alternatives and possible modifications of the 
quantum measure for non-Markovianity studied here. One possibility is to employ 
alternative distance measures for quantum states under which trace preserving 
quantum operations are contractive, such as the relative entropy or the 
Bures distance which is based on the fidelity \cite{Hayashi}. In particular, the 
relative entropy represents a possibility which is natural both from an information 
theoretic perspective and from the point of view of nonequilibrium 
thermodynamics, since the negative rate of change of the entropy relative
to an invariant thermal equilibrium state can be 
interpreted as entropy production \cite{Wehrl,Spohn}. However, a disadvantage of 
the relative entropy is given by the fact that it is often infinite, leading to 
singularities of the measure \cite{BLP}. In several cases, in particular for infinite 
dimensional Hilbert spaces, the determination of the trace distance could be an 
extremely difficult task. An analytical formula for the trace distance is not even 
known for Gaussian quantum states. It seems that in those cases it is much 
easier to work with the Bures distance or the fidelity which also leads to useful 
lower and upper bounds for the trace distance \cite{Vasile}. The Hilbert-Schmidt 
distance which is technically much easier to deal with cannot be used for the 
quantification of non-Markovianity because trace preserving quantum operations 
are, in general, not contractive for this metric \cite{Wang}. A further possibility is to 
define the measure for quantum non-Markovianity by means of alternative 
functionals of the family of dynamical maps which quantify the dynamical increase 
of the chosen distance measure for quantum states. 

Both the quantum measure for non-Markovianity and the witness for initial
system-environment correlations studied here have been demonstrated 
very recently to be experimentally measurable quantities 
\cite{NatPhys,EPL,Hefei,Milan}. These experiments
have paved the way for a series of further investigations of quantum memory 
effects in composite, multipartite open systems. A central goal of the quantum 
theory of open system is thus the design of appropriate schemes and models for 
their theoretical treatment and, in particular, the development of efficient 
numerical methods for the determination of quantum measures for 
non-Markovianity.

\begin{acknowledgments}
I would like to thank Jyrki Piilo and Bassano Vacchini for fruitful collaboration
and many interesting discussions on the diverse aspects of non-Markovian 
quantum dynamics. Thanks also to Chuan-Feng Li, Bi-Heng Liu, Elsi-Marie Laine, 
Sabrina Maniscalco, Govinda Clos, Andrea Smirne, Stefan Fischer, Manuel 
Gessner and Steffen Wissmann. Financial support by the German Academic 
Exchange Service (DAAD) is gratefully acknowledged.
\end{acknowledgments}

\end{document}